\documentclass{article}
\usepackage{graphicx}  
\usepackage{amsmath}
\usepackage{subcaption}
\usepackage{amssymb}   
\usepackage{bm} 
\usepackage{dcolumn}
\usepackage{color}
\usepackage{mathrsfs}

\usepackage{indentfirst}
\usepackage{relsize}

\usepackage{amsfonts}
\usepackage{varioref}
\usepackage[margin=1.3in]{geometry}
\usepackage{slashed}
\RequirePackage[colorlinks,citecolor=blue,urlcolor=magenta,linkcolor=blue]{hyperref}
\usepackage{comment}
\def\p{\partial}
\def\e{\epsilon}

\def\be{\begin{equation}}
\def\ee{\end{equation}}

\labelformat{section}{Section #1} 
\labelformat{subsection}{Section #1} 
\labelformat{subsubsection}{Section #1}
\labelformat{subsubsubsection}{Section #1}
\labelformat{equation}{Eq.~(#1)} 
\labelformat{figure}{Fig.~#1} 
\labelformat{subfigure}{Fig.~\thefigure#1} 
\labelformat{table}{Tab.~#1} 
\labelformat{appendix}{Appendix #1}

\title{\bf  $\xi R \phi^2$ coupling, cosmological constant and quantum gravitational correction to Newton's potential}
\author{Avijit Sen Majumder\footnote{senmajumderavijit@gmail.com}\, \, and \, Sourav Bhattacharya\footnote{sbhatta.physics@jadavpuruniversity.in}\\ 
\small{\it Relativity and Cosmology Research Centre, Department of Physics, Jadavpur University, Kolkata 700 032, India}\\}
\begin{document}
\maketitle
\begin{abstract}
\noindent
This letter investigates the contribution of the $\sqrt{-g}\xi R\phi^2$ non-minimal interaction to the long range gravitational potential for  massive scalar fields, found from the non-relativistic limit of the  2-2 scattering amplitude with graviton exchanges. Such coupling  is most naturally motivated from the renormalisation of a scalar field theory with quartic self interaction in a curved spacetime. This  is qualitatively different from the minimal ones like $ \sqrt{G} h^{\mu\nu}T_{\mu\nu}$, as the vertices corresponding to the former does not explicitly contain any scalar momenta, but  instead explicitly contains the momentum carried by   graviton line. For the minimal vertex, the  long range gravitational potential up to  one loop $({\cal O}(G), {\cal O}(G^2))$ was obtained earlier from the terms non-analytic in the transfer momentum, $q^{-2},\ q^{-1},\ \ln q^2 $, yielding potentials respectively like $r^{-1}$, $r^{-2}$, $r^{-3}$. However owing to the aforesaid  explicit appearance of graviton's transfer momentum  for the non-minimal vertices, the leading contribution  in this case comes at ${\cal O}(\xi G^2)$, and turns out  to be subleading compared to even $r^{-3}$.  In order to complement  this `screening' effect, we consider the three graviton vertex generated by the $\sim \Lambda \sqrt{-g}/G$ term in the action, where $\Lambda$ is the cosmological constant. This vertex does not contain   any graviton momentum explicitly.  With this vertex, and assuming short scale scattering much small compared to the Hubble horizon, we compute the seagull, the vacuum polarisation and the fish diagrams and obtain the   2-2 scattering amplitudes. The leading two body gravitational potential  at ${\cal O}(\xi \Lambda G^2 )$ behaves like $ r^{-1}$, even though it is much subleading compared to  Newton's  potential due to the appearance of $\Lambda$. We also discuss the scenario  where this potential  dominates the aforesaid  ${\cal O}(\xi G^2)$ one.
\end{abstract}
%



\section{Introduction}\label{S1}

\noindent
There has been a lot of effort in the past and present to understand the low energy effects of perturbative quantum gravity, even though it is not renormalisable. The goal behind this is to see its predictions at the first few orders of the perturbation theory, with the hope that they will be verifiable in a not too far away future. We also hope that such  perturbative framework will be smoothly embedded in a complete theory of quantum gravity one day. In particular, the perturbative quantum gravity might tell us whether at all the concept of graviton as a {\it quantum} massless spin-2 field is meaningful. We refer our reader to~\cite{Donoghue:1993eb, Muzinich:1995uj, Hamber:1995cq, Bjerrum-Bohr:2002gqz, Bjerrum-Bohr:2002fji, Burgess:2003jk, Reuter:2004nv, Kirilin:2006en, Holstein:2008sx, Anber:2011ut, Rodigast:2009zj, Marunovic:2012pr, Bjerrum-Bohr:2015vda, Battista:2017xlm, deBrito:2020wmp, Akhoury:2013yua, Saltas:2016nkg, Saltas:2016awg, Latosh:2020jyq, Latosh:2022ydd} and references therein for various tree and one loop computations in this framework,  chiefly focussing on the leading quantum corrections to the long range  gravitational potential, emerging from the graviton exchanges between two non-relativistic particles. See~\cite{vanDam:1970vg, Bjerrum-Bohr:2014zsa, Jusufi:2016sym} and references therein for computation of bending of light. We refer our reader to e.g.~\cite{Goldberger:2004jt, Foffa:2016rgu, Levi:2018nxp, Wu:2023jwd} and references therein for the computations on the corrections to the Newton potential via effective field theory techniques. See also~\cite{Cheung:2018wkq}  for a technique for extracting  potential from the four point amplitude in the post Minkowski expansion. Moreover, see~\cite{Toms:2008dq, Toms:2009zz, Toms:2009vd, Toms:2010vy, Toms:2011zza} for inclusion of a cosmological constant at subhorizon scales, where the background metric can still be taken to be $\eta_{\mu\nu}$. Finally, see e.g.~\cite{Perez-Nadal:2008byr, McDonald:2015iwt, Frob:2016fcr, Frob:2017smg, Boran:2017fsx} and references therein for discussions in curved spacetimes like the de Sitter. We note that the long range predictions of  perturbative quantum gravity can be interpreted as an effective field theory framework making well defined physical predictions, irrespective  of its ultraviolet completion. This serves as one of the chief motivations behind engaging into such low energy computations associated with Einstein's or other alternative gravity theories.

In this work we wish to probe the curvature-scalar non-minimal coupling term,  $\xi \sqrt{-g} R\phi^2$, \ref{nm1}, in the context of  long range gravitational potential  for  massive scalar fields, where $\xi$ is the non-minimal coupling parameter.  Such coupling naturally arises in the presence of background curvature while renormalising a scalar field theory with quartic self-interaction~\cite{Parker:2009uva, Buchbinder}. What will be the contribution of this term at the lowest order in  $\xi$, at tree and one loop level to the correction of the Newton potential?  This coupling is qualitatively different from the minimal ones. This is because for a scalar field theory with quadratic kinetic term, in the matter-graviton interaction there can be no derivative of the graviton (such as $\kappa h^{\mu\nu}T_{\mu\nu}$), whereas for the non-minimal interaction, there is no derivative of the scalar at all (\ref{nm8'a}).   For $\xi=0$, the leading gravitational potential from  2-2 scalar scattering amplitudes  up to one loop  was computed earlier from the terms non-analytic in the transfer momentum carried by  gravitons ($q^{-2},\ q^{-1},\ \ln q^2 $), yielding results  up to ${\cal O}(r^{-3})$, e.g.~\cite{Bjerrum-Bohr:2002gqz}. For $\xi \neq 0$ however,  we will see below that replacement of any minimal matter-graviton vertex by a non-minimal vertex results in  the appearance of graviton's transfer momentum in the numerator of the corresponding scattering amplitudes. Consequently, no long range contribution  is found at tree level. The leading such contribution is found at ${\cal O}(\xi G^2)$ for some one loop diagrams,   and they are  subleading  compared to even  $r^{-3}$.   For the two scalar-one graviton and two scalar-two graviton vertices generated by the $\sqrt{-g}\xi R\phi^2 $ term, the diagrams will have the same topology as that of $\xi=0$~\cite{Hamber:1995cq, Bjerrum-Bohr:2002gqz}.  Thus in order to understand whether there exists any stronger contribution, we must check if there is any way to complement this `screening' effect. We will see that this is possible in the Einstein gravity,  if we consider a cosmological constant and the corresponding ${\cal O}(\Lambda \kappa)$  three graviton vertex generated by the $-4\Lambda \sqrt{-g}/\kappa^2$ term, \ref{nm1}. Unlike the usual three graviton vertex generated by  $2\sqrt{-g} R/\kappa^2$, there will be no derivative of the graviton here. Accordingly, we will see in \ref{1loop} that the seagull (\ref{f1}), the vacuum polarisation (\ref{f2}) and the fish diagrams (\ref{f3}) will contribute to the gravitational potential, at ${\cal O}(\xi \Lambda G^2)$, with the leading behaviour of $r^{-1}$. We will further emphasise qualitative differences of the potential thus found with the standard ones ($\xi=0=\Lambda$). We also recall that for a conformally symmetric scalar field ($m^2=0,\ \xi =1/6$),  static spherically symmetric black hole solutions are known to exist since long~\cite{Bocharova:1970skc, Bekenstein:1974sf, Martinez:2002ru, Bhattacharya:2013hvm}. To the best of our knowledge, the $\xi \sqrt{-g} R\phi^2$  coupling, even though qualitatively different from standard minimal ones,  is relatively less investigated from  quantum gravity perspective. We refer our reader for example,  to~\cite{Moss:2014nya, Shapiro:2015ova, Oda:2015sma, Saltas:2015vsc} for discussion on renormalisation in curved spacetime, asymptotic safety and quantum gravity, and Higgs inflation in the presence of such  coupling.  Due to its aforementioned qualitatively different nature from the standard matter-graviton coupling, it seems to be an important job to further probe the effect of the first on quantum gravity predictions.

Now,  in general the presence of a $\Lambda$ prevents us from doing quantum field theory in a flat background, and instead one should do it in a de Sitter background, which reads, for example, in the static coordinatisation,
\be
ds^2= - fdt^2 +f^{-1}dr^2 +r^2 d\Omega^2
\label{dS0}
\ee
where $f=1-\Lambda r^2/3$.
However much inside the Hubble horizon ($r \ll \sqrt{3/\Lambda}$), the spacetime can be split into a Minkowski background plus a perturbation.   The Schwarzschild-de Sitter spacetime, describing the gravitational filed of a static mass inside the de Sitter universe, is given by $f(r)=1-2MG/r-\Lambda r^2/3$ in \ref{dS0}.
The region with smaller $r$ values are dominated by the attraction due to the mass $M$, and the larger ones by the repulsion due to $\Lambda$. The attraction decreases monotonically with increasing  $r$, whereas the repulsion increases. Accordingly, there is a scale,
$R_{\rm TA, max}= ( 3MG/\Lambda)^{1/3},$
known as the maximum turn around radius, around which the attraction and repulsion balance each other, e.g.~\cite{Bhattacharya:2017yix}. It is clear that we may do a decomposition of flat spacetime plus perturbation inside this radius, essentially to look into {\it short wavelength} phenomena. Such decomposition in the presence of a $\Lambda$ has been done earlier in~\cite{Toms:2008dq, Toms:2009zz, Toms:2009vd, Toms:2010vy, Toms:2011zza}.     One may reasonably  anticipate that these short scale  effects  will be tiny while compared  to the $\Lambda=0$ cases hitherto investigated. Nevertheless, we believe probing such effects are important, in order to further understand the nature of the matter-gravity interaction.  We also note that in the primordial  inflationary  scenario and in the infrared regime, one can encounter non-perturbative secular large logarithms while computing the quantum corrections to the two body gravitational potential, e.g.~\cite{Frob:2016fcr} and references therein. Understanding the role of the $\xi \sqrt{-g}R \phi^2$ coupling in the inflationary and large scale context can be an important task, which we reserve for the future. 

We wish to emphasise that the phrases {\it short scale/short wavelength} appearing anywhere in this letter is  to be understood as the {\it subhorizon/subHubble} scale. Even though the potential is long range, it  is valid only up to the  scale of the maximum turn around radius of a structure. Beyond this scale the expansion of the universe dominates, and the Hubble flow will essentially invalidate the notion of a potential as an explicit time independent non-relativistic quantity. 

In this regard,  we also note that  while the maximum turn around radius mentioned above can  be thought of  as a local test of cosmological constant/dark energy~\cite{Bhattacharya:2017yix}, the other effects of $\Lambda$ on the gravitational physics inside the maximum turn around radius are expected to be small compared to that  of the mass of the gravitating object.  $\Lambda$ dominates at very large scales beyond the maximum turn around radius, and contributes to phenomena  like the gravitational redshift, or correlations originating from the  very early universe, etc..  Nevertheless,  with increasing measurement facilities such as the event horizon telescope, one can hope to  probe the effect of $\Lambda$ in short scale physics (like the solar system) as well, in the not too far away future, like the perihelion precession, bending of light and the Shapiro delay. This might give us new insight about the nature of the dark energy. Thus attempts to understand such phenomena's perturbative quantum gravity corrections are also well motivated.    

Before we go into the computations, we will summarise below the basic technical ingredients we will need. We will work with the  mostly positive signature of the metric in four spacetime dimensions. Our notation for  symmetrisation  will be : $X_{(\mu\nu)}= X_{\mu\nu}+X_{\nu\mu}$.

\section{The basic ingredients}
\noindent Our action reads
\begin{eqnarray}
S= \frac{2}{\kappa^2}\int d^4 x \sqrt{-g}  \ \left( R- 2\Lambda\right)-\frac12\int  d^4 x \sqrt{-g}\ \left[ g^{\mu\nu} (\p_{\mu}\phi) (\p_{\nu}\phi)+ m^2 \phi^2 + \xi R \phi^2\right]   
\label{nm1}
\end{eqnarray}
where $\kappa^2 = 32 \pi G$. We wish to treat the non-minimal interaction term perturbatively. The metric is broken into background plus perturbation
\begin{eqnarray}
g_{\mu\nu} =\eta_{\mu\nu} + \kappa h_{\mu\nu}; \qquad g^{\mu\nu}= \eta^{\mu\nu}-\kappa h^{\mu\nu}+ \kappa^2 h^{\mu}{}_{\alpha}h^{\alpha\nu}+\cdots  
\label{nm3}
\end{eqnarray}
Using  the De Donder gauge, 
$$\p_{\mu}\left(h^{\mu}{}_{\nu}-\frac12 \delta^{\mu}_{\nu} h \right)=0 $$
the leading, free and linearised Einstein equation reads
\be
\p^2 \left(h_{\mu\nu}-\frac12 h \eta_{\mu\nu} \right)- \frac{2\Lambda}{\kappa} \eta_{\mu\nu}=0
\label{qg27}
\ee
giving
\be
h_{\mu\nu}= h_{\mu\nu}\vert_{\rm hom} - \frac{\Lambda r^2}{3\kappa} \eta_{\mu\nu},
\label{qg28}
\ee
where the first term on the right-hand side is the usual homogeneous massless spin-2 plane wave solution, which  will give the  field quantisation, whereas the second term corresponds to the linearised, classical de Sitter perturbation yielding the background curvature, $R=4\Lambda$. The quantisation  for $h_{\mu\nu}\vert_{\rm hom}$ yields the graviton propagator. The ground state for the theory will be taken to be the standard Poincarr\'e invariant Minkowski vacuum, subject to the flat  background, \ref{nm3}. Note that in the presence of  $\Lambda$, this cannot be possible at large scales, due to particle creation and the subsequent instability of the initial  vacuum state. This happens if we are concerned with length scales comparable to or higher than  the Hubble horizon. Such large scale, infrared effects are essentially non-perturbative and non-equilibrium, so that the standard in-out $S$-matrix formalism of quantum field theory which we will be using here, fails. Scattering has no well defined meaning in such dynamical backgrounds and one needs to do the in-in formalism to compute expectation values or correlation functions, e.g.~\cite{Boran:2017fsx} and references therein. This problem is not relevant to our present scenario, for as we have stated in the preceding section, we are concerned here only up to the length scale of a structure (like a galaxy) defined by its maximum turn around radius. This length scale is much subhorizon and hence we may safely ignore the aforementioned large scale IR effects for our present purpose. This means that the $\Lambda$ term appearing in \ref{qg28} is much small compared to unity. We will assert this scenario in our calculations in the following  by taking the Compton wavelength of the scattering  particles to be much small compared to the horizon length. Also, since the current observed value of $\Lambda$ is tiny, ${\cal O} (10^{-52}\ {\rm m}^{-2})$, we will restrict our computations  to ${\cal O}(\Lambda)$ only.    

We now make the following leading expansions for the scalar and the $\Lambda$ part of the action,
\begin{equation}
    \begin{split}
        S_{\phi} =&-\frac12\int d^4 x \left[ (\p \phi)^2 +m^2 \phi^2\right] + \frac{\kappa}{2}\int d^4 x h^{\mu\nu}\left[(\p_{\mu}\phi)(\p_{\nu}\phi)-\frac12 \eta_{\mu\nu}\left( (\p \phi)^2+m^2 \phi^2\right) \right]\\ &
        -\frac{\kappa^2}{2} \int d^4 x \left[ \left(h^{\mu}{}_{\alpha} h^{\alpha\nu}-\frac12 h h^{\mu\nu}+ \frac{h^2}{8}\eta^{\mu\nu}-\frac14 h_{\alpha \beta} h^{\alpha \beta}\eta^{\mu\nu}\right)(\p_{\mu}\phi)(\p_{\nu}\phi) + \frac{m^2}{4}\left( \frac{h^2}{2}-h_{\mu\nu}h^{\mu\nu}\right)\phi^2\right]\\ & +\frac{\xi}{2}\int d^4 x \left[ \kappa \phi^2\p^2 h - \kappa^2 \phi^2 \left\{- \frac14 h \p^2 h - \frac12 (\p^{\lambda}h^{\mu\alpha})(\p_{\alpha}h_{\mu\lambda})  +  h^{\mu\nu} \p^2 h_{\mu\nu} +\frac34(\p^{\lambda} h^{\mu\nu})(\p_{\lambda}h_{\mu\nu})\right\}\right] + \ {\cal O}(\kappa^3, \xi \kappa^3) \\ 
        S_{\Lambda} =& -\frac{4\Lambda}{\kappa^2}\int d^d x \left[1+ \frac{\kappa h}{2} + \frac{\kappa^2}{4}\left( \frac{h^2}{2} -  h_{\mu\nu}h^{\mu\nu}\right)+\frac{\kappa^3}{2}\left(\frac13 h_{\mu\nu}h^{\nu\lambda}h_{\lambda}{}^{\mu}-\frac14 h h_{\mu\nu}h^{\mu\nu}+ \frac{h^3}{24} \right)  \right] +\ {\cal O}(\Lambda \kappa^2)
        \label{nm8'a}
    \end{split}
\end{equation}
where we have used $m^2  \gg \Lambda$, i.e., the Compton wavelength of the scalar is much small compared to the Hubble horizon, as stated above. Note in \ref{nm8'a} that there is a term quadratic in $h_{\mu\nu}$ in $S_{\Lambda}$, which could also make a contribution proportional  to $\Lambda h_{\mu\nu}$ in \ref{qg27}.  However, this is subleading compared to the $\eta_{\mu\nu}$ term, and hence has been ignored.  Also as  \ref{qg28} suggests,  both quantum $h_{\mu\nu}\vert_{\rm hom}$ and the classical $\Lambda r^2$ terms  make contributions in the above. However, we will only be concerned about the quantum effects generated by $h_{\mu\nu}\vert_{\rm hom}$, as the classical terms cannot contribute in forming a vertex or in carrying momenta. Moreover in  $S_{\Lambda}$, the classical $\Lambda r^2$ term will make contributions  beyond linear order in $\Lambda$, hence  are further subleading in our subhorizon scenario. Thus    we  have a three graviton vertex of ${\cal O}(\kappa \Lambda)$ from $S_{\Lambda}$.  There is no explicit appearance of the graviton momentum here, for it contains no derivative of the graviton. This vertex will serve the key role in our following computations.

The scalar and the graviton propagators read
\begin{eqnarray}
&& \Delta (k)=-\frac{1}{k^2 +m^2}; \qquad  \Delta_{\mu\nu\alpha\beta} (k)= -\frac{ P_{\mu\nu\alpha\beta}}{k^2}; \qquad P_{\mu\nu\alpha\beta}=\frac12 \left( \eta_{\mu\alpha}\eta_{\nu\beta}+ \eta_{\mu\beta}\eta_{\nu\alpha}-\eta_{\mu\nu}\eta_{\alpha\beta}\right)
\label{nm6}
\end{eqnarray}
where $\mu,\nu$ and $\alpha, \beta$ pairwise correspond to two graviton fields, $h_{\mu\nu}$ and $h_{\alpha\beta}$.  
 
 The usual three graviton vertex originating from the $\sqrt{-g}R/2\kappa^2$ term (with momenta $l$, $q$, $l-q$ along the three lines) reads, e.g.~\cite{Bjerrum-Bohr:2002gqz},
\begin{equation}
    \begin{split}
        V^{(3)\mu\nu}_{\alpha\beta \gamma\delta}(\kappa;l, q)=& -\frac{i\kappa}{2} \left[  P_{\alpha\beta \gamma\delta}\left(l^{\mu}l^{\nu}+(l-q)^{\mu}(l-q)^{\nu}+q^{\mu}q^{\nu} -\frac32 \eta^{\mu\nu}q^2\right) \right. \\&\left.+2 q_{\lambda} q_{\sigma} \left(I_{\alpha\beta}{}^{\sigma\lambda}I_{\gamma\delta}{}^{\mu\nu}+I_{\gamma\delta}{}^{\sigma\lambda}I_{\alpha\beta}{}^{\mu\nu}-I_{\alpha\beta}{}^{\mu\sigma}I_{\gamma\delta}{}^{\nu\lambda}- I_{\gamma\delta}{}^{\mu\sigma}I_{\alpha\beta}{}^{\nu\lambda}  \right) \right. \\&\left.+\left\{q_{\lambda} q^{\mu}\left(\eta_{\alpha\beta} I_{\gamma\delta}{}^{\nu\lambda}+\eta_{\gamma\delta} I_{\alpha\beta}{}^{\nu\lambda} \right)   + q_{\lambda} q^{\nu}\left(\eta_{\alpha\beta} I_{\gamma\delta}{}^{\mu\lambda}+\eta_{\gamma\delta} I_{\alpha\beta}{}^{\mu\lambda} \right) - q^2 \left(\eta_{\alpha\beta} I_{\gamma\delta}{}^{\mu\nu}+\eta_{\gamma\delta} I_{\alpha\beta}{}^{\mu\nu}\right)  \right.\right. \\&\left. \left. - \eta^{\mu\nu}q_{\sigma}q_{\lambda}\left(\eta_{\alpha \beta}I_{\gamma\delta}{}^{\sigma\lambda}+ \eta_{\gamma\delta}I_{\alpha \beta}{}^{\sigma\lambda}\right)  \right\}\right.\\&\left.+ \left\{-2q_{\lambda} \left(I_{\alpha\beta}{}^{\lambda \sigma} I_{\gamma\delta \sigma}{}^{\nu} (l-q)^{\mu}+I_{\alpha\beta}{}^{\lambda \sigma} I_{\gamma\delta \sigma}{}^{\mu} (l-q)^{\nu} +  I_{\gamma\delta}{}^{\lambda \sigma}I_{\alpha\beta\sigma}{}^{\nu}l^{\mu}+ I_{\gamma\delta}{}^{\lambda \sigma}I_{\alpha\beta\sigma}{}^{\mu}l^{\nu} \right) \right.\right. \\&\left.\left. +q^2 \left(I_{\alpha\beta\sigma}{}^{\mu}I_{\gamma\delta}{}^{\nu\sigma}+ I_{\gamma\delta\sigma}{}^{\mu}I_{\alpha\beta}{}^{\nu\sigma} \right)+\eta^{\mu\nu}q_{\sigma}q_{\lambda}\left(I_{\alpha\beta}{}^{\lambda\rho}I_{\gamma\delta \rho}{}^{\sigma}+ I_{\gamma\delta}{}^{\lambda\rho}I_{\alpha\beta\rho}{}^{\sigma} \right)  \right\} \right.\\&\left. 
        + \left\{\left(l^2+(l-q)^2 \right) \left(I_{\alpha\beta}{}^{\mu\sigma}I_{\gamma\delta\sigma}{}^{\nu} + I_{\gamma\delta}{}^{\mu\sigma}I_{\alpha\beta\sigma}{}^{\nu}  -\frac12 \eta^{\mu\nu}{\cal P}_{\alpha\beta\gamma\delta}\right)- \left(I_{\gamma\delta}{}^{\mu\nu}\eta_{\alpha\beta}l^2+I_{\alpha\beta}{}^{\mu\nu}\eta_{\gamma\delta}(l-q)^2  \right)  \right\} \right],
        \label{qg25}
    \end{split}
\end{equation}
where
$$I_{\mu\nu\lambda \rho}= \frac12 \left(\eta_{\mu\lambda}\eta_{\nu\rho}+\eta_{\mu\rho}\eta_{\nu\lambda} \right) $$
We will use the standard integral~\cite{Peskin:1995ev}
\be
I^{r}_s= \int \frac{d^d k}{(2\pi)^d} \frac{(k^2)^r}{(k^2-a^2)^s}= \frac{i(-1)^{r+s}}{(4\pi)^{d/2}}\frac{1}{(a^2)^{s-r-d/2}}\frac{\Gamma(r+d/2)\Gamma(s-r-d/2)}{\Gamma(d/2)\Gamma(s)},
\label{nm8'}
\ee
as well as the Fourier transforms
\begin{eqnarray}
\int \frac{d^3 \vec{q}}{(2\pi)^3}\frac{e^{i\vec{q}\cdot \vec{r}}}{\vec{q}^2}=\frac{1}{4\pi r},\qquad \int \frac{d^3 \vec{q}}{(2\pi)^3}\frac{e^{i\vec{q}\cdot \vec{r}}}{|\vec{q}|}=\frac{1}{2\pi^2 r^2}, \qquad \int \frac{d^3 \vec{q}}{(2\pi)^3}e^{i\vec{q}\cdot \vec{r}}\ln \vec{q}^2=-\frac{1}{2\pi r^3}
\label{nm8add}
\end{eqnarray}
in the following calculations. Also, we will take the dummy virtual momentum named $l^{\mu}$ appearing below  to run from $-\infty$ to $+\infty$. The amplitudes will be evaluated in the usual relativistic framework, whereas the non-relativistic limit (such as $k_1\cdot k_2 \approx -m_1m_2$) will be  taken  after the evaluation.

\begin{figure}[htbp]
    \centering
    \includegraphics[width=0.36\linewidth]{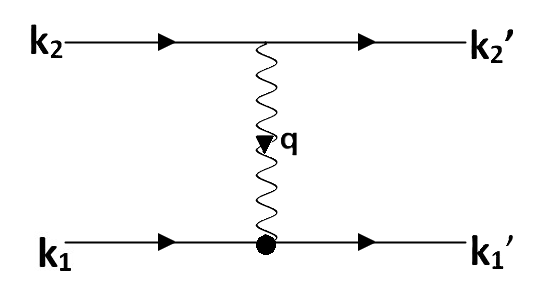}
    \hspace{2cm}
    \includegraphics[width=0.36\linewidth]{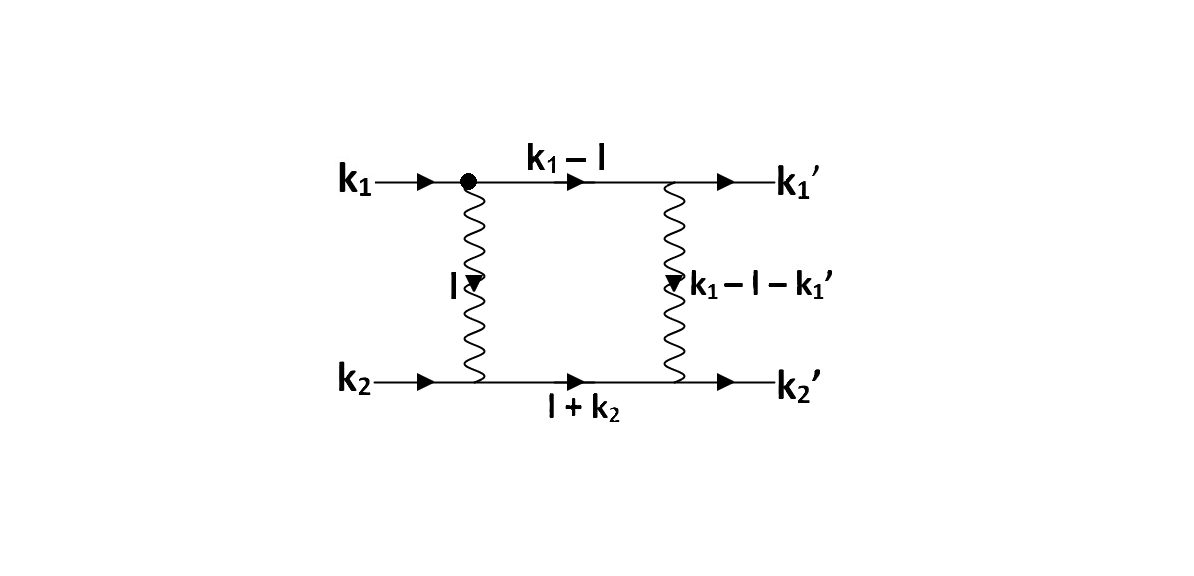}
    \caption{\small \it The tree and the box diagrams for 2-2 scattering with graviton exchanges. Solid and wavy lines respectively denote scalar and graviton propagators. The thick circle denotes a non-minimal vertex of ${\cal O}(\xi \kappa)$. The external momenta are on shell, ${k^2_1}={k'}_1^2=-m_1^2$, ${k^2_2}={k'}_2^2=-m_2^2$. Both diagrams yield no non-analytic contribution in the transfer momentum, $q=k_1-k'_1= k'_2-k_2$, and hence to the long range gravitational potential. The leading contribution behaves as $ r^{-4}$ at ${\cal{O}}(\xi G^2)$ from some other diagrams.  See the main text for details.}
    \label{f0}
\end{figure}

Let us first see how the non-minimal term contributes to the gravitational interaction, without any $\Lambda$. The tree level process is given by the first of  \ref{f0}. From \ref{nm8'a}, we see that the non-minimal one graviton-two scalar vertex function reads: $i\xi \kappa (-q^2)\eta_{\mu\nu}/2$. The $k_2-k'_2$ vertex is taken to be the usual minimal one graviton-two scalar vertex. The Feynman amplitude reads,
\be
- i{\cal M} = -i. i\kappa \frac{i (-q^2) \xi \kappa \eta_{\mu\nu}P^{\mu\nu\alpha \beta}\left( k_{1(\alpha}k'_{1\beta)}-\eta_{\alpha\beta}(k_1\cdot k'_1 +m_1^2)\right)}{2^2q^2} 
\label{tree}
\ee
where $q= k_1-k'_1=k'_2-k_2$. The long range gravitational potential is defined in the non-relativistic limit by the Fourier transform of ${\cal M}/4m_1m_2$ in the first Born approximation, originating from the functions non-analytic in  $q^2$. However, the potential corresponding \ref{tree} is proportional to $\delta^3(\vec{r})$, i.e., not a {\it long range} one we are interested in. It can be checked explicitly that analogous things occurs for the box diagram of \ref{f0}. As a further explicit example, let us consider the triangle diagram generated by pinching the $k_1-l$-line of the box diagram. This will generate a two scalar-two graviton vertex in the upper half of the diagram. Let us first imagine that this vertex is non-minimal, produced by~\ref{nm8'a}.  It is not difficult to see that this diagram generates a momentum integral of the kind
\be
\sim \int\frac{d^d l}{(2\pi)^d}\frac{ l\cdot (l-q) \ k_2\cdot (l-k'_2)\ k'_2 \cdot (l-k'_2)}{l^2 (l-q)^2\left( (l-k'_2)^2 + m_2^2\right)},
\label{refadd}
\ee
which gives a leading  potential $\sim G^2 \xi m_2^2/m_1 r^4$ in the non-relativistic limit. Note the appearance of the mass of the scalar in the denominator. Even though long range, this potential is much subleading and  we wish to look for leading results than this. Similar result is found if instead we pinch the $(l+k_2)$-line of the second of \ref{f0}.  If on the other hand, we take either the $k_2$ or the $k'_2$ vertex to be non-minimal, we encounter integrals like
$$\sim \int\frac{d^d l}{(2\pi)^d}\frac{ l^2  \ k_2\cdot (l+k_2)}{l^2 (l-q)^2\left( (l+k_2)^2 + m_2^2\right)}\quad  {\rm Or,} \quad \sim \int\frac{d^d l}{(2\pi)^d}\frac{ (l-q)^2  \ k'_2\cdot (l+k_2)}{l^2 (l-q)^2\left( (l+k_2)^2 + m_2^2\right)},$$
both of which are independent of $q$ and hence makes no contribution to the long range potential. Likewise, for the other one loop diagrams as well, we find either they do not yield any potential at all, or their leading contribution is ${\cal O}(r^{-4})$.

As we have already emphasised,  evidently the above situation arises due to the transfer momentum carried by  the graviton appearing in the numerator. This  motivates us to bring in the $\Lambda$-three graviton vertex mentioned above,  and to see whether we may obtain contribution leading compared to $r^{-4}$.  There can be three relevant diagrams involving this vertex at one loop, which we wish to evaluate below.   \ref{nm8'} will be one of the chief ingredients for the following calculations. 
%

\section{The $\Lambda$-three graviton vertex and the gravitational potential at ${\cal O}(\xi \kappa^4 \Lambda)$}\label{1loop}
%
\begin{figure}[htbp]
    \centering
    \includegraphics[width=0.42\linewidth]{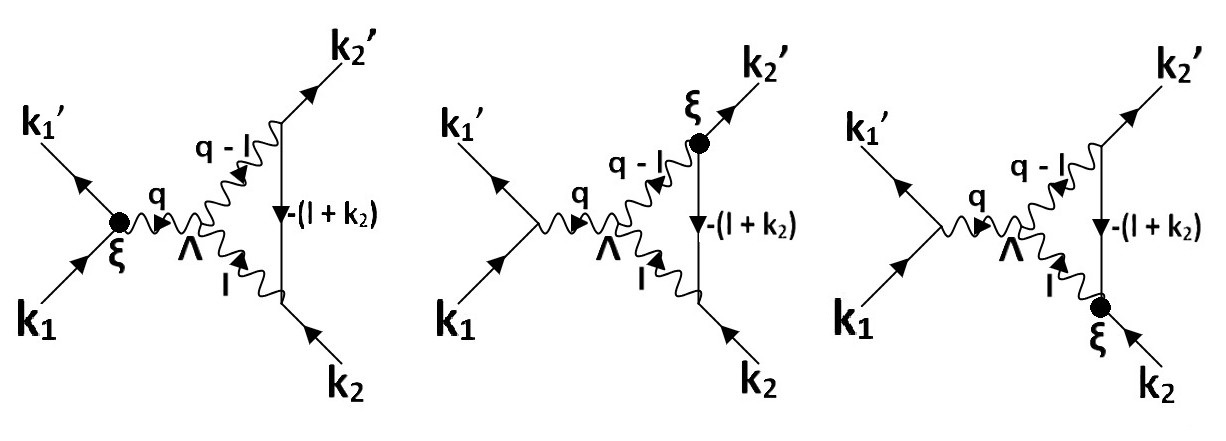}
    \caption{\small \it  The seagull diagram for 2-2 scalar scattering at ${\cal O}(\xi \kappa^4 \Lambda)$, with the $\Lambda$-three graviton vertex.   There are three more diagrams, found by the interchange, $(k_1, k'_1) \ \leftrightarrow \ (k_2,k'_2) $.
} 
    \label{f1}
\end{figure}
\begin{figure}[htbp]
    \centering
    \includegraphics[width=0.42\linewidth]{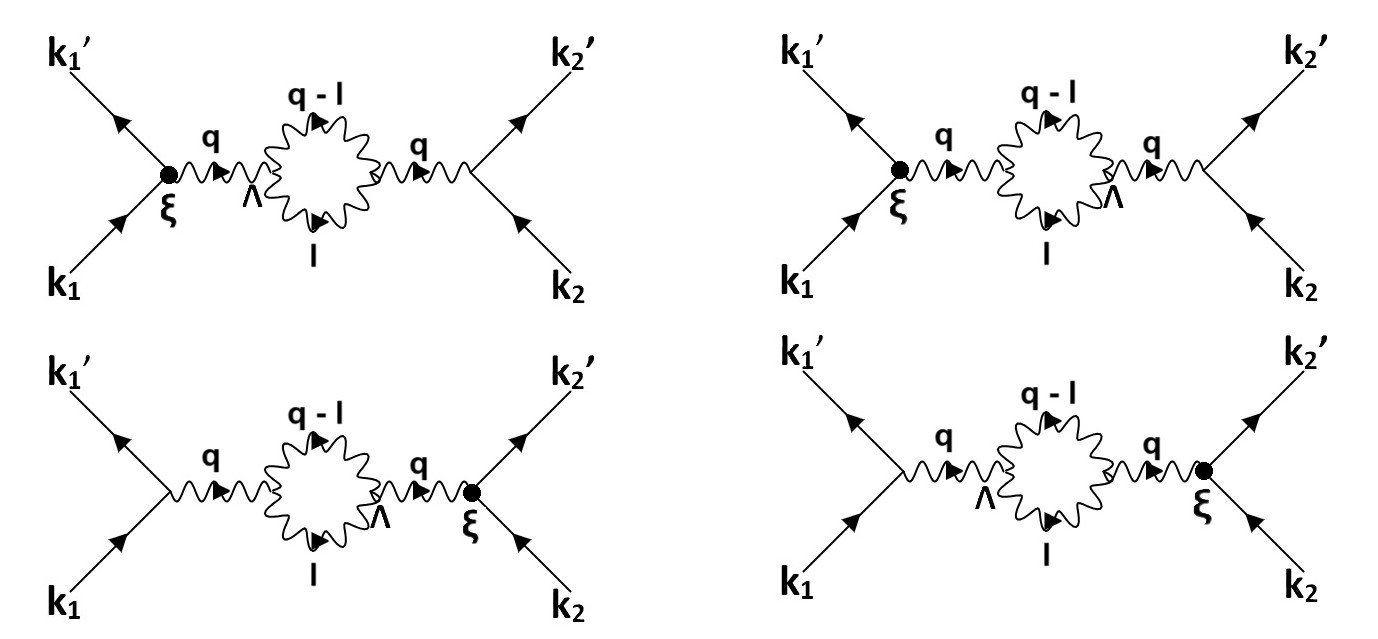}
    \caption{\small \it The vacuum polarisation diagram for the 2-2 scalar scattering at ${\cal O}(\xi \kappa^4 \Lambda)$. One of the cubic graviton vertices corresponds to the $\Lambda$-three graviton vertex, \ref{nm8'a}, whereas the other one is the usual, \ref{qg25}. Also, as earlier, the thick circle denotes a non-minimal vertex. }
    \label{f2}
\end{figure}
\begin{figure}[htbp]
    \centering
    \includegraphics[width=0.42\linewidth]{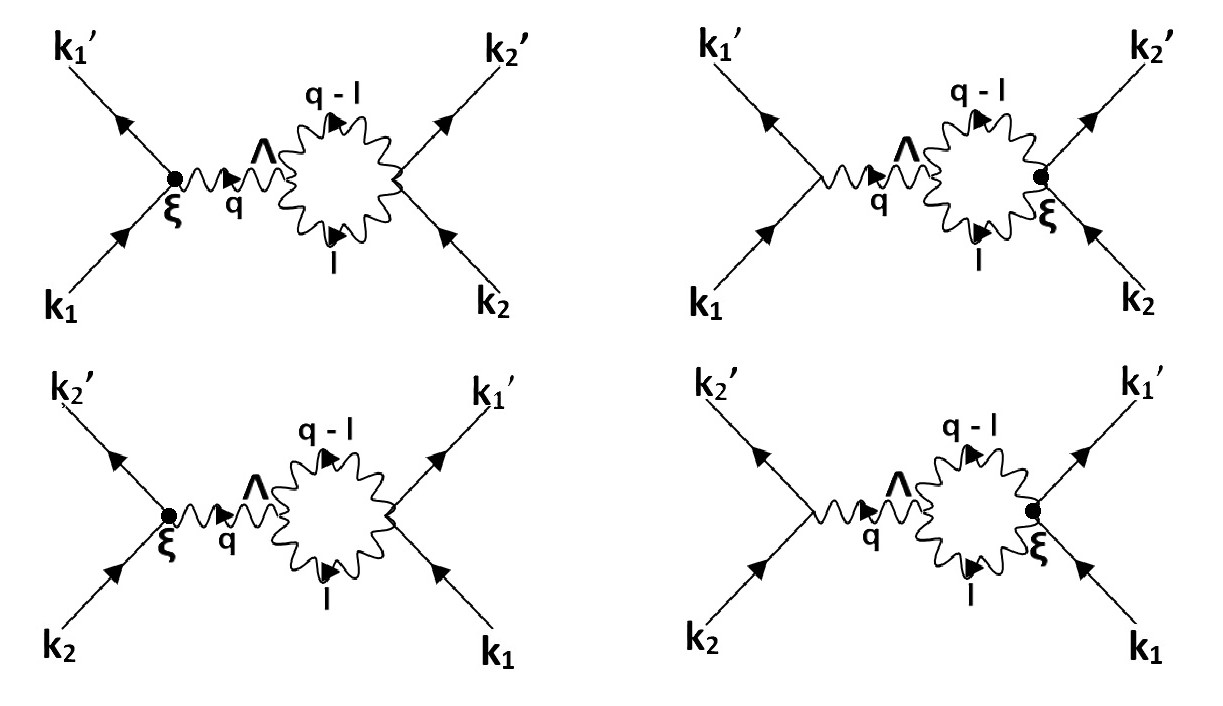}
    \caption{\small \it The fish diagram for the 2-2 scalar scattering at ${\cal O}(\xi \kappa^4 \Lambda)$. The cubic vertex is the $\Lambda$-three graviton vertex. }
    \label{f3}
\end{figure}
\subsection{The seagull}
There are total six diagrams for the seagull class, \ref{f1}. Using \ref{nm8'a} for $S_{\Lambda}$, the first of them gives the contribution ($d=4-\e$) 
{\small
\begin{equation}
    \begin{split}
      &  -i {\cal M}^{\rm 2.1} = \frac{\mu^{4\e}\Lambda \kappa^4 \xi}{2^3}\int \frac{d^d l}{(2\pi)^d} q^2\eta^{\mu\nu}P_{\mu\nu\alpha \beta}\eta^{\alpha\beta}\eta_{\lambda\rho}\eta_{\gamma\delta}P^{\lambda\rho\lambda'\rho'}P^{\gamma \delta \gamma' \delta'} \left\{k'_{2(\lambda'} (l-k'_2)_{\rho')}- \eta_{\lambda'\rho'}(k_2'\cdot (l-k_2')-m_2^2)\right\} \\ & \times \frac{\left\{k_{2(\gamma'} (l-k_2)_{\delta')}- \eta_{\gamma'\delta'}(k_2\cdot (l-k_2)-m_2^2)\right\}}{q^2l^2(l-q)^2((l-k_2')^2+m_2^2)}\\ &
        - \frac{\mu^{4\e}\Lambda \kappa^4 \xi}{2^2}\int \frac{d^d l}{(2\pi)^d} q^2\eta^{\mu\nu}P_{\mu\nu\alpha \beta}\eta^{\alpha\beta}\eta_{\lambda'\mu'}\eta_{\nu'\rho'}P^{\lambda'\rho'\alpha'\beta'}P^{\mu'\nu'\gamma' \delta'} \left\{k'_{2(\gamma'} (l-k'_2)_{\delta')}- \eta_{\gamma'\delta'}(k'_2\cdot (l-k_2')-m_2^2)\right\} \\ & \times \frac{\left\{k_{2(\alpha'} (l-k_2)_{\beta')}- \eta_{\alpha'\beta'}(k_2\cdot (l-k_2)-m_2^2)\right\}}{q^2l^2(l-q)^2((l-k_2')^2+m_2^2)}\\ &
        - \frac{\mu^{4\e}\Lambda \kappa^4 \xi}{2}\int \frac{d^d l}{(2\pi)^d} q^2\eta^{\mu\nu} \eta^{\mu'\lambda'}\eta^{\nu'\rho'} \eta_{\alpha \beta}        P_{\mu\nu\mu'\nu'} P^{\alpha\beta \gamma\delta}P_{\lambda'\rho'}{}^ {\gamma' \delta'} \left\{k'_{2(\gamma'} (l-k'_2)_{\delta')}- \eta_{\gamma'\delta'}(k'_2\cdot (l-k_2')-m_2^2)\right\} \\ & \times \frac{\left\{k_{2(\gamma} (l-k_2)_{\delta)}- \eta_{\gamma\delta}(k_2\cdot (l-k_2)-m_2^2)\right\}}{q^2l^2(l-q)^2((l-k_2')^2+m_2^2)}\\ & 
        +\mu^{4\e}\Lambda \kappa^4 \xi\int \frac{d^d l}{(2\pi)^d} q^2\eta^{\alpha\beta} P_{\alpha\beta \mu\nu} P^{\nu\lambda \gamma\delta} P_{\lambda}{}^{\mu\alpha'\beta'}\left\{k'_{2(\gamma} (l-k'_2)_{\delta)}- \eta_{\gamma\delta}(k'_2\cdot (l-k_2')-m_2^2)\right\}  \frac{\left\{k_{2(\alpha'} (l-k_2)_{\beta')}- \eta_{\alpha'\beta'}(k_2\cdot (l-k_2)-m_2^2)\right\}}{q^2l^2(l-q)^2((l-k_2')^2+m_2^2)}\\
        =& -2\Lambda \kappa^4 \xi \mu^{4\e}\int \frac{d^d l}{(2\pi)^d} \frac{k_2\cdot (l-k_2)k'_2\cdot (l-k'_2) - 2m_2^2 k_2'\cdot (l-k_2')- 2m_2^2 k_2\cdot (l-k_2) +4m_2^4}{l^2(l-q)^2((l-k_2')^2+m_2^2)}\\& + 2\Lambda \kappa^4 \xi \mu^{4\e}\int \frac{d^d l}{(2\pi)^d} \frac{k_2\cdot k_2'(l-k_2')\cdot(l-k_2)+ k_2'\cdot (l-k_2)k_2\cdot(l-k_2')-m_2^2k_2'\cdot (l-k_2')-m_2^2k_2\cdot (l-k_2)+2m_2^4}{l^2(l-q)^2((l-k_2')^2+m_2^2)}\\&
        +2\Lambda \kappa^4 \xi \mu^{4\e}\int \frac{d^d l}{(2\pi)^d} \frac{k_2\cdot (l-k_2)k_2'\cdot (l-k_2')-2m_2^2k_2'\cdot (l-k_2')-2m_2^2k_2\cdot (l-k_2)+4m_2^4}{l^2(l-q)^2((l-k_2')^2+m_2^2)}\\ &
        -2\Lambda \kappa^4 \xi \mu^{4\e} \int \frac{d^d l}{(2\pi)^d}\frac{k_2\cdot k_2'(l-k_2')\cdot(l-k_2)+ k_2'\cdot (l-k_2)k_2\cdot(l-k_2')-m_2^2k_2'\cdot (l-k_2')-m_2^2k_2\cdot (l-k_2)+2m_2^4}{l^2(l-q)^2((l-k_2')^2+m_2^2)}=0, 
        \label{qg29}
    \end{split}
\end{equation}}
where $\mu$ is the mass scale for renormalisation.  Likewise, if we interchange pairwise $k_1, k'_1$ with $k_2, k'_2$, we obtain a vanishing contribution. \\

We next place the non-minimal vertex on the $k_2'$ momentum, the second of \ref{f1}. Using the on-shell and non-relativistic relations ($k_1^2={k'_1}^2  = -m_1^2$, $k_2^2 ={k'_2}^2 = -m_2^2$, $k_1\cdot k_2 \approx -m_1 m_2,\, k'_1\cdot k'_2 \approx -m_1 m_2 $), we have
\begin{flalign*}
         -i {\cal M}^{\rm 2.2} =& \frac{\mu^{4\e}\Lambda \kappa^4 \xi}{2^4 q^2}k_{1(\mu} k'_{1\nu)} 
        \eta^{\alpha \beta}\eta^{\gamma\delta}\eta^{\lambda \rho}\eta^{\lambda' \rho'}P^{\mu\nu}{}_{\alpha\beta}P_{\lambda \rho\lambda' \rho'}P_{\gamma\delta}{}^{\alpha' \beta'}\int \frac{d^d l}{(2\pi)^d}\frac{k_{2(\alpha'}(l+k_2)_{\beta')}- l \cdot k_2 \ \eta_{\alpha'\beta'}}{l^2((l+k_2)^2+m_2^2)} &&
        \\ &
        + \frac{\mu^{4\e}\Lambda \kappa^4 \xi}{2 q^2}k_{1(\mu} k'_{1\nu)}P^{\mu\nu}{}_{\mu'\nu'}\eta^{\gamma\delta}P^{\nu'\lambda'}{}_{\gamma\delta}P_{\lambda'}{}^{\mu'\alpha\beta} \int \frac{d^d l}{(2\pi)^d}\frac{k_{2(\alpha}(l+k_2)_{\beta)}- l \cdot k_2 \ \eta_{\alpha\beta}}{l^2((l+k_2)^2+m_2^2)}&&
        \\&
        -\frac{\mu^{4\e}\Lambda \kappa^4 \xi}{2^3 q^2}k_{1(\alpha} k'_{1\beta)}\eta^{\lambda'\rho'}P^{\alpha\beta}{}_{\lambda'\rho'}\eta^{\lambda\rho}P_{\mu\nu\lambda \rho}P^{\mu\nu\gamma\delta} \int \frac{d^d l}{(2\pi)^d}\frac{k_{2(\gamma}(l+k_2)_{\delta)}- l \cdot k_2 \ \eta_{\gamma\delta}}{l^2((l+k_2)^2+m_2^2)}&&
        \\&
        -\frac{\mu^{4\e}\Lambda \kappa^4 \xi}{2^2 q^2}k_{1(\alpha} k'_{1\beta)}P^{\alpha\beta}{}_{\mu\nu}\eta^{\lambda\rho}P_{\lambda \rho\lambda' \rho'}\eta^{\lambda'\rho'}P^{\mu\nu\gamma\delta} \int \frac{d^d l}{(2\pi)^d}\frac{k_{2(\gamma}(l+k_2)_{\delta)}- l \cdot k_2 \ \eta_{\gamma\delta}}{l^2((l+k_2)^2+m_2^2)}  && 
\end{flalign*}
\small{\begin{equation*}
    \begin{split}
        =&  -\frac{\mu^{4\e}\Lambda \kappa^4 \xi m_1^2}{2 q^2} \int \frac{d^d l}{(2\pi)^d}\left[\frac{2m_2^2}{l^2((l+k_2)^2+m_2^2)}+\frac{1}{l^2}- \frac{1}{(l+k_2)^2+m_2^2}\right]
        +\frac{\mu^{4\e}\Lambda \kappa^4 \xi}{2 q^2} \int \frac{d^d l}{(2\pi)^d} \frac{2l\cdot \left(m_1m_2 k_1'+ m_1m_2 k_1- m_1^2 k_2 \right)-4m_1^2m_2^2}{l^2((l+k_2)^2+m_2^2)}\\ &
         +\frac{\mu^{4\e}\Lambda \kappa^4 \xi m_1^2}{4 q^2} \int \frac{d^d l}{(2\pi)^d}\left[\frac{2m_2^2}{l^2((l+k_2)^2+m_2^2)}+\frac{1}{l^2}- \frac{1}{(l+k_2)^2+m_2^2}\right]
        -\frac{\mu^{4\e}\Lambda \kappa^4 \xi}{ q^2} \int \frac{d^d l}{(2\pi)^d} \frac{2l\cdot \left(m_1m_2 k_1'+ m_1m_2 k_1- m_1^2 k_2 \right)-4m_1^2m_2^2}{l^2((l+k_2)^2+m_2^2)}  
    \end{split}
\end{equation*}}
\begin{equation}
    \begin{split}
        = &-\frac{\mu^{4\e}\Lambda \kappa^4 \xi m_1^2}{4 q^2} \int \frac{d^d l}{(2\pi)^d}\left[\frac{2m_2^2}{l^2((l+k_2)^2+m_2^2)}- \frac{1}{(l+k_2)^2+m_2^2}\right]
        -\frac{\mu^{4\e}\Lambda \kappa^4 \xi}{ 2q^2} \int \frac{d^d l}{(2\pi)^d} \frac{2l\cdot \left(m_1m_2 k_1'+ m_1m_2 k_1- m_1^2 k_2 \right)-4m_1^2m_2^2}{l^2((l+k_2)^2+m_2^2)}
        \label{qg30}
    \end{split}
\end{equation}
which we evaluate using \ref{nm8'}.
The divergence can be absorbed in the renormalisation counterterm for the one graviton-two scalar vertex diagram embedded in \ref{f1}. The part relevant to the potential (i.e., non-analytic in the transfer momentum $q^2$) reads
\begin{eqnarray}
&&-i {\cal M}^{\rm 2.2}\equiv -\frac{i\Lambda \kappa^4\xi m_1^2 m_2^2}{8\pi^2q^2}
\label{qg31}
\end{eqnarray}
If we place the non-minimal vertex on the $k_2$ vertex, we obtain exactly the same result as above. Likewise, if we pairwise interchange the $k_1, k_1'$ and $k_2, k_2'$ vertices, we obtain the same result. Thus, we have the total contribution to the gravitational potential for the seagull class of diagrams, \ref{f1}, 
\begin{eqnarray}
&&-i {\cal M}^{\rm seagull}_{\rm tot.}\equiv -\frac{i\Lambda \kappa^4\xi m_1^2 m_2^2}{2\pi^2q^2}
\label{qg32}
\end{eqnarray}
 By \ref{nm8add}, the above amplitude will give an $r^{-1}$ potential in the non-relativistic limit. Note that there is no subleading contribution to the potential from the seagull diagrams. This is simply because only the second and the third diagrams of \ref{f2} (along with the pairwise interchange of the $k_1, k_1'$ and $k_2, k_2'$ vertices) contribute to the above potential. Due to the placement of the nonminimal vertex in these diagrams we have $(l-q)^2$ or $l^2$ appearing in the numerator of the corresponding scattering amplitudes. This results in the factorisation of $q^{-2}$ and momentum integral independent of the transfer momentum $q$, \ref{qg30}. Accordingly, the contribution in \ref{qg32} is exact, and there is no  other terms (like $q^{-1}$ or $\ln q^2$) which can make subleading contributions to the potential.

\subsection{The vacuum polarisation}
The vacuum polarisation class of diagrams are given by~\ref{f2}. There are two three graviton vertices involved here -- one is associated with  $\Lambda$, whereas the other one is the usual, \ref{qg25}.
Using $k_1\cdot q \approx q^2/2$, $k_1'\cdot q \approx -q^2/2$, $k_2\cdot q \approx -q^2/2$ and $k'_2\cdot q \approx q^2/2$, we compute for the first diagram
\begin{equation}
    \begin{split}
        -i {\cal M}^{\rm 3.1} =& \frac{3\mu^{4\e}\kappa^4\Lambda \xi}{8}(-q^2)\frac{\eta^{\mu'\nu'}}{q^2}P_{\mu'\nu'\alpha'\beta'}\eta^{\alpha'\beta'}\eta^{\lambda'\rho'}\eta^{\gamma'\delta'}P_{\lambda'\rho'\lambda\rho}P_{\gamma'\delta'\gamma\delta}\\& \times\left\{k_{2(\mu}k'_{2\nu)}- \eta_{\mu\nu}(k_2\cdot k_2'+m^2_2) \right\}\int \frac{d^d l}{(2\pi)^d}\frac{\bar{\tau}^{\mu''\nu''\lambda\rho\gamma\delta}}{l^2 (l-q)^2}\frac{1}{q^2}P_{\mu''\nu''}{}^{\mu\nu}
        +3\mu^{4\e}\kappa^4\Lambda \xi (-q^2)\frac{\eta^{\mu'\nu'}}{q^2}P_{\mu'\nu'\alpha'\beta'}P^{\beta'\gamma'}{}_{\lambda\rho}P_{\gamma'}{}^{\alpha'}{}_{\gamma\delta}\\ &
        \times\left\{k_{2(\mu}k'_{2\nu)}- \eta_{\mu\nu}(k_2\cdot k_2'+m^2_2) \right\}\int \frac{d^d l}{(2\pi)^d}\frac{\bar{\tau}^{\mu''\nu''\lambda\rho\gamma\delta}}{l^2 (l-q)^2}\frac{1}{q^2}P_{\mu''\nu''}{}^{\mu\nu}
        -\frac{3\mu^{4\e}\kappa^4\Lambda \xi}{4}(-q^2)\frac{\eta^{\mu'\nu'}}{q^2}P_{\mu'\nu'\alpha'\beta'}\eta^{\alpha'\beta'} P_{\lambda'\rho'\lambda\rho}P^{\lambda'\rho'}{}_{\gamma\delta}\\ &
        \times\left\{k_{2(\mu}k'_{2\nu)}- \eta_{\mu\nu}(k_2\cdot k_2'+m^2_2) \right\}\int \frac{d^d l}{(2\pi)^d}\frac{\bar{\tau}^{\mu''\nu''\lambda\rho\gamma\delta}}{l^2 (l-q)^2}\frac{1}{q^2}P_{\mu''\nu''}{}^{\mu\nu}
        -\frac{3\mu^{4\e}\kappa^4\Lambda \xi}{2}(-q^2)\frac{\eta^{\mu'\nu'}}{q^2}P_{\mu'\nu'\alpha'\beta'}P^{\alpha'\beta'}{}_{\lambda\rho}\eta_{\lambda'\rho'}P^{\lambda'\rho'}{}_{\gamma\delta}\\ & \times\left\{k_{2(\mu}k'_{2\nu)}- \eta_{\mu\nu}(k_2\cdot k_2'+m^2_2) \right\}\int \frac{d^d l}{(2\pi)^d}\frac{\bar{\tau}^{\mu''\nu''\lambda\rho\gamma\delta}}{l^2 (l-q)^2}\frac{1}{q^2}P_{\mu''\nu''}{}^{\mu\nu}=0
        \label{qg33}
    \end{split}
\end{equation}
where $\bar{\tau}^{\mu\nu\lambda\rho\gamma\delta}$ is the three graviton vertex of \ref{qg25}, modulo the factor $-i\kappa/2$.
Likewise, if the $k_2-k_2'$ is non-minimal and it is connected to the $\Lambda$-three graviton vertex (i.e., the third of~\ref{f2}), the contribution vanishes.  Similarly, we find that all the contributions for this class of diagrams vanish.

\subsection{The fish diagram}

This class of diagrams is given by \ref{f3}, and the cubic vertex here is the $\Lambda$-three graviton vertex. Let us first place the non-minimal vertex in the $k_1-k_1'$ vertex. The two scalar-two graviton vertex is the usual (${\cal O}(\kappa)$). We thus have for the first diagram,
\begin{equation}
    \begin{split}
        -i {\cal M}^{\rm 4.1}=& - \Lambda \mu^{4\e}\kappa^4 \xi (-q)^2\eta^{\mu\nu}\frac{P_{\mu\nu\lambda' \rho'}}{q^2} \eta^{\lambda' \rho'}\int \frac{d^dl}{(2\pi)^d}\frac{\eta_{\alpha'\beta'}P^{\alpha'\beta'}{}_{\lambda\eta} \eta_{\rho'\sigma'}P^{\rho'\sigma'}{}_{\rho\sigma}}{l^2(l-q)^2} \times \bar{\tau}_{2,2}^{\lambda\eta\rho\sigma} \\ &
        - 8\mu^{4\e}\Lambda \kappa^4\xi (-q)^2\eta^{\mu\nu}\frac{P_{\mu\nu\mu' \nu'}}{q^2} P^{\nu'\lambda'}{}_{\lambda\eta}P_{\lambda'}{}^{\mu'}_{\rho\sigma}\int \frac{d^dl}{(2\pi)^d}\frac{1}{l^2(l-q)^2}\times \bar{\tau}_{2,2}^{\lambda\eta\rho\sigma} 
        \label{qg34}
    \end{split}
\end{equation}
\begin{equation*}
    \begin{split}
        &+ 2\mu^{4\e}\Lambda \kappa^4\xi (-q)^2\eta^{\mu\nu}\frac{P_{\mu\nu\mu' \nu'}}{q^2}\eta^{\mu'\nu'} P^{\lambda'\rho'}{}_{\lambda\eta} P_{\lambda'\rho'\rho\sigma} \int \frac{d^dl}{(2\pi)^d}\frac{1}{l^2(l-q)^2}\times \bar{\tau}_{2,2}^{\lambda\eta\rho\sigma} \\ &
        + 4\mu^{4\e}\Lambda \kappa^4\xi (-q)^2\eta^{\mu\nu}\frac{P_{\mu\nu\mu' \nu'}}{q^2}P^{\mu'\nu'}{}_{\lambda\eta}\eta^{\lambda'\rho'} P_{\lambda'\rho'\rho\sigma} \int \frac{d^dl}{(2\pi)^d}\frac{1}{l^2(l-q)^2}\times \bar{\tau}_{2,2}^{\lambda\eta\rho\sigma}
    \end{split}
\end{equation*}
where we have abbreviated for the two scalar-two graviton vertex~\cite{Bjerrum-Bohr:2002gqz},
\be
\bar{\tau}_{2,2}^{\lambda\eta\rho\sigma} ={\tau}_{2,2}^{\lambda\eta\rho\sigma}/\kappa^2 =
\left[\left\{I^{\eta\lambda\alpha \delta}I^{\rho\sigma \beta}{}_{\delta}-\frac14 \left(\eta^{\eta\lambda}I^{\rho\sigma\alpha\beta}+\eta^{\rho\sigma}I^{\eta\lambda\alpha\beta} \right)      \right\}k_{2(\alpha}k'_{2\beta)} -\frac12\left(I^{\eta\lambda\rho\sigma}-\frac12 \eta^{\eta \lambda}\eta^{\rho\sigma}\right)(k_2\cdot k'_2+m_2^2) \right]
\label{qg35}
\ee
A little algebra shows that \ref{qg34} is vanishing. Likewise, if the ${\cal O}(\kappa)$ non-minimal vertex is $k_2-k'_2$, we find a vanishing contribution (the third of \ref{f3}).

\noindent
We next come to the case when the 2-scalar 2-graviton vertex is non-minimal, ${\cal O}(\xi \kappa^2)$. We have for the second of  \ref{f3}, 
\begin{equation}
    \begin{split}
        -i {\cal M}^{\rm 4.2}=&  \frac{\mu^{3\e}\Lambda \kappa^4 \xi}{4} k_{1(\mu}k'_{1\nu)} P^{\mu\nu\mu'\nu'}\eta_{\mu'\nu'} \int \frac{d^dl}{(2\pi)^d}\frac{1}{l^2(l-q)^2q^2}\\ &
        \times \left\{-\frac12 \eta^{\lambda'\rho'}P_{\lambda'\rho'}{}^{\mu\alpha} \eta^{\alpha'\beta'}P_{\alpha'\beta'\mu\lambda}l^{\lambda}(l-q)_{\alpha}+\frac34 l\cdot(l-q)\eta^{\lambda'\rho'}P_{\lambda'\rho'}{}^{\mu\nu}\eta^{\alpha'\beta'}P_{\alpha'\beta'\mu\nu}\right\}\\ &
+2\mu^{3\e}\Lambda \kappa^4\xi k_{1(\mu'}k'_{1\nu')}P^{\mu'\nu'}{}_{\lambda'\rho'}\int \frac{d^dl}{(2\pi)^d}\frac{1}{l^2(l-q)^2 q^2}\left\{-\frac12 P^{\rho'\gamma'\mu\alpha}{P_{\gamma'}{}^{\lambda'}}_{\mu\lambda}l^{\lambda}(l-q)_{\alpha} +\frac34 l\cdot(l-q) P^{\gamma'\rho'\mu\nu}P^{\lambda'}{}_{\gamma'\mu\nu}\right\}
 \\ &
-\mu^{3\e}\Lambda \kappa^4\xi k_{1(\mu'}k'_{1\nu')} P^{\mu'\nu'\alpha'\beta'}\int \frac{d^dl}{(2\pi)^d}\frac{1}{l^2(l-q)^2 q^2}
 \left\{-\frac12 P_{\alpha'\beta'}{}^{\mu\alpha}\eta^{\lambda'\rho'}P_{\lambda'\rho'\mu\lambda} l^{\lambda}(l-q)_{\alpha}+\frac34 l\cdot(l-q)P_{\alpha'\beta'}{}^{\mu\nu}\eta^{\lambda'\rho'}P_{\lambda'\rho'\mu\nu}\right\}\\ &
-\mu^{3\e}\frac{\Lambda \kappa^4\xi}{2} k_{1(\mu'}k'_{1\nu')} P^{\mu'\nu'\alpha'\beta'}\eta_{\alpha'\beta'} \int \frac{d^dl}{(2\pi)^d}\frac{1}{l^2(l-q)^2 q^2} \left\{-\frac12 P_{\gamma\delta}{}^{\mu\alpha}P^{\gamma\delta}{}_{\mu\lambda} l^{\lambda}(l-q)_{\alpha}+\frac34 l\cdot(l-q)P_{\gamma\delta}{}^{\mu\nu}P^{\gamma\delta}{}_{\mu\nu}\right\} \\ 
%
=&\frac{\mu^{3\e}m_1^2 \Lambda \kappa^4 \xi}{4} \int \frac{d^d l}{(2\pi)^d}\frac{1}{l^2 (l-q)^2}-\frac{\mu^{3\e}\Lambda \kappa^4 \xi}{ q^2}\int \frac{d^d l }{(2\pi)^d }\frac{2k_{1\mu} k'_{1\nu} l^{\mu}l^{\nu}+ q^2 l\cdot q/2}{l^2 (l-q)^2} 
    \label{qg34'}
    \end{split}
\end{equation}
Only the first integral contributes to the potential at leading order, which is $\sim \ln q^2$. Likewise, by replacing $m_1$ by $m_2$, we obtain the result if $k_2-k'_2$  is the non-minimal two scalar-two graviton vertex. Putting things together now, including \ref{qg32}, we have the total Feynman amplitude for 2-2 scattering at ${\cal O}(\xi \Lambda \kappa^4)$, in the non-relativistic limit,
\begin{eqnarray}
&&-i {\cal M}^{\rm tot., non-analytic}\vert_{\rm NR} = -\frac{i\Lambda \kappa^4\xi m_1^2 m_2^2}{2\pi^2\vec{q}^2}-\frac{i(m_1^2+m_2^2) \Lambda \kappa^4 \xi}{4\pi^2} \ln \frac{\vec{q}^2}{\mu^2}  + \ {\cal O} \left(\vec{q}^2 \ln \frac{\vec{q}^2}{\mu^2}\right).
\label{qg35'}
\end{eqnarray}
Note that since $|\vec q|$ is non-relativistic, we must have $|\vec q|/m \ll 1$. We expect the contribution from  the first  term in the potential to dominate the other.

Before we compute the potential corresponding to \ref{qg35'}, a couple of comments on  vanishing of some Feynman amplitudes are in order. Specifically, we have seen that two of the seagull diagrams falling into the topology of the first of \ref{f1} vanish. Likewise  the amplitudes for all the vacuum polarisation diagrams,~\ref{f2}, vanish, whereas for~\ref{f3}, only two of the diagrams give non-vanishing contributions. Such vanishing amplitudes might correspond to some symmetry associated with the non-minimal and/or the $\Lambda$-three graviton vertices.   For all these cases, note that the diagrams are nicely symmetric with respect to the position of the non-minimal vertex. We also note that two of the diagrams~\ref{f3}, even though they look symmetric with respect to the location of the non-minimal vertex, the said vertex is two scalar-two graviton and not two scalar-one graviton, and the diagrams yield non-vanishing contributions, \ref{qg34'}. Thus  the aforesaid symmetry, if there is any, seems to be  only relevant to the non-minimal matter-graviton 3-point vertex, along  with the $\Lambda$-three graviton vertex.  In particular, from the structure of \ref{qg33} it is not difficult to see that if we compute the one loop self energy of the graviton at ${\Lambda \kappa^2}$ due to one $\Lambda$-three graviton vertex and one usual three graviton vertex, the self energy would vanish. This corresponds to the cancellation between various terms arising due to the $\Lambda$-vertex. However, any specific nature or form of this speculated symmetry remains  elusive to us so far.  In this regard, we also refer our reader to e.g.~\cite{Seery:2007we}  for demonstration of vanishing of one loop one and two point functions in a theory of inflation containing higher derivative terms for the inflaton. We also refer to~\cite{Ivanov:2022qqt} for vanishing of some scattering amplitudes off the Kerr black hole.

\section{The final result and discussion}
\noindent
The non-relativistic gravitational potential in the first Born approximation is defined as~\cite{Hamber:1995cq, Bjerrum-Bohr:2002gqz},
$$V(\vec{r})= - \frac{1}{4m_1m_2}\int \frac{d^3 \vec{q}}{(2\pi)^3} e^{i \vec{q}\cdot \vec{r}} {\cal M}^{\rm tot., non-analytic}\vert_{\rm NR}(\vec{q}^2)$$
Using the above, we obtain  from \ref{qg35'}, \ref{nm8add}, 
 the leading contribution of the non-minimal coupling parameter $\xi$ to the two body long range gravitational potential,
\begin{eqnarray}
V(G^2, \Lambda, \xi,r)= -\frac{32G^2 \Lambda \xi m_1 m_2}{\pi r} \left[ 1- \frac{m_1^2+m_2^2}{m^2_1 m^2_2}\frac{1}{ r^2}\right]
\label{qg36'}
\end{eqnarray}

We note the appearance of  masses in the denominator of the second term on the right hand side. Each mass has to be understood as the inverse of the corresponding Compton wavelength. Thus the first term dominates the second. Perhaps a two loop computation for the class of diagrams that gave the first term would produce a result comparable to the second. Note also that this second  term has originated from the Fourier transform of $\ln \vec{q}^2$, and any such logarithm  originates from the  term like $(q^2/\mu^2)^{\e/2}\cdot \e^{-1}$  coming from \ref{nm8'}. Hence it is essentially a quantum contribution, whereas the first is not.

In this regard, we note  the  $\xi=0$ part of the gravitational potential up to one loop~\cite{Hamber:1995cq, Bjerrum-Bohr:2002gqz},
\be
V(\Lambda =0, r)\vert_{\rm minimal}= - \frac{G m_1 m_2}{r}\left[1- \frac{G(m_1+m_2)}{r} -\frac{127 G }{30 \pi^2 r^2} \right]
\label{ref2}
\ee
In \ref{qg36'}, we do not have any term of first order in $G$ at all, due to the absence of tree level contribution. In fact the vanishing tree level long range contribution with $\xi\neq 0$ actually {\it saves} $\xi$, for otherwise a contribution like $\xi Gm_1m_2/r$ would  have constrained $\xi$ very severely from simple solar system data, as is the case for the parameter of the Brans-Dicke theory~\cite{Clifton:2011jh}. Second, we also note that one can use the $\Lambda$-3 graviton vertex used here to evaluate the two body potential with $\xi=0$. However, the result is expected to be subleading compared to the $\Lambda=0$ ones at our subhorizon scale.\\

\noindent

To the best of our knowledge, the vertex associated with $\Lambda$ has not been considered earlier.   Note also that the potential of \ref{qg36'}  contains two fundamental length scales --- the shortest or Planck length corresponding to $G$, and the longest or Hubble horizon scale corresponding to $\Lambda$. We believe this is a novel feature encoded in physics of length scale much small compared to that of the horizon. \\

\noindent
Let us now compare the leading term of \ref{qg36'} with that of the ${\cal O}(G^2\xi)$ term with $\Lambda=0$ (${\cal O}(r^{-4})$) discussed below \ref{refadd}.  The ratio of the first to the second  for $m_2 \gg m_1$ behaves as,
$$\sim \Lambda r^2\cdot  \frac{m_1}{m_2} \cdot  m_1 r $$
Recall that a mass  term is understood as its associated inverse Compton wavelength ($mc/h$).  Imagine now that $m_2, m_1$ are respectively  the sun and the Mercury. Taking $\Lambda \sim 10^{-52}{\rm m}^{-2}$ and  $r\sim 10^{10} {\rm m}$ to be the radius of the perihellion,  the above ratio becomes $\sim 10^{36}$. It is clear that this ratio will keep increasing with increasing $r$ or $m_1$ or both.  Taking now $m_1$ to be the electron, the ratio is instead found to be $10^{-72}$. Thus for `large' $m_1$ values, the first term of \ref{ref2} would dominate over the ${\cal O}(G^2\xi)$ terms. We believe the dominating role of a term associated with $\Lambda$ compared to the $r^{-4}$ term for large mass objects, even in the much subhorizon solar scale physics,  is interesting in its own right.  \\

Let us now compare \ref{qg36'} with the standard Newton potential for two massive scalars.
The leading parts of~\ref{qg36'} and the above has a ratio of ${\cal O}(G\Lambda \xi)\sim \xi (L_P/L_C)^2$, where $L_C \sim \Lambda^{-1/2}$ is the size of the cosmological event horizon and $L_P$ is the Planck length.  Thus this ratio can at most be $10^{-10}$ in the context of primordial inflationary universe and much lower for our current universe. Also, the ratio of the leading part of \ref{qg36'}  and the most subleading part of \ref{ref2} reads $\sim \Lambda \xi r^2$, which is also small in the subhorizon scale we are interested in. We also note that the dependence on $m_1$ and $m_2$ for the two potentials, \ref{qg36'}, \ref{ref2}, are qualitatively very different. This originates from the fact that   \ref{qg36'}  contains $\Lambda$, which has mass dimension two. Also  as we have mentioned towards the end of \ref{S1}, perturbative quantum gravity predictions in the flat background with a $\Lambda$ are expected to be subleading compared to that of the $\Lambda=0$ case, at least in our current universe where $\Lambda$ is very tiny.  Such computations in the early inflationary universe might yield some even non-perturbative, phenomenologically significant contributions, e.g.~\cite{Frob:2016fcr}. Note that in such background, computation of scattering at large scales is not possible due to vacuum instability. Instead the effective action formalism might be useful. We wish to come back to the issue of $\xi R \phi^2$ coupling and gravitational potential in the context of early inflationary universe in our future work.

There are further things which seem to need our attention in the context of the matter-graviton non-minimal coupling. For example, a complete evaluation of the two body potential at various post-Minkowski order would be interesting. An investigation on the effect of $\xi$ on gravitational radiation seems to be an important task. These might lead to important constraint on $\xi$, giving us valuable insight about the matter-gravity interaction.  It would also be interesting to investigate the effect of $\xi$ on self energy, various matter-graviton vertex functions, and RG perspectives.    We wish to address these issues in our future publications. \\

\noindent
{\bf \large Acknowledgements :} SB would like to thank Amitabha Lahiri and Parthasarathi Majumdar for useful discussions. The authors are indebted to anonymous referees for careful and critical reading of the manuscript and for making many useful comments and suggestions.  



\begin{thebibliography}{99} 



\bibitem{Donoghue:1993eb}
J.~F.~Donoghue,
{\it Leading quantum correction to the Newtonian potential},
Phys. Rev. Lett.\textbf{72}, 2996-2999 (1994)
[arXiv:gr-qc/9310024 [gr-qc]].

\bibitem{Muzinich:1995uj}
I.~J.~Muzinich and S.~Vokos,
{\it Long range forces in quantum gravity},
Phys. Rev. D\textbf{52}, 3472 (1995)
[arXiv:hep-th/9501083 [hep-th]].


  
  \bibitem{Hamber:1995cq}
H.~W.~Hamber and S.~Liu,
{\it On the quantum corrections to the Newtonian potential},
Phys. Lett. B\textbf{357}, 51-56 (1995)
[arXiv:hep-th/9505182 [hep-th]].


\bibitem{Bjerrum-Bohr:2002gqz}
N.~E.~J.~Bjerrum-Bohr, J.~F.~Donoghue and B.~R.~Holstein,
{\it Quantum gravitational corrections to the nonrelativistic scattering potential of two masses},
Phys. Rev. D\textbf{67}, 084033 (2003)
[erratum: Phys. Rev. D\textbf{71}, 069903 (2005)]
[arXiv:hep-th/0211072 [hep-th]].

\bibitem{Bjerrum-Bohr:2002fji}
N.~E.~J.~Bjerrum-Bohr, J.~F.~Donoghue and B.~R.~Holstein,
{\it Quantum corrections to the Schwarzschild and Kerr metrics},
Phys. Rev. D\textbf{68}, 084005 (2003)
[erratum: Phys. Rev. D\textbf{71}, 069904 (2005)]
[arXiv:hep-th/0211071 [hep-th]].

\bibitem{Burgess:2003jk}
C.~P.~Burgess,
{\it Quantum gravity in everyday life: General relativity as an effective field theory},
Living Rev. Rel.\textbf{7}, 5-56 (2004)
[arXiv:gr-qc/0311082 [gr-qc]].

\bibitem{Reuter:2004nv}
M.~Reuter and H.~Weyer,
{\it Running Newton constant, improved gravitational actions, and galaxy rotation curves},
Phys. Rev. D\textbf{70}, 124028 (2004)
[arXiv:hep-th/0410117 [hep-th]].


\bibitem{Kirilin:2006en}
G.~G.~Kirilin,
{\it Quantum corrections to the Schwarzschild metric and reparametrization transformations},
Phys. Rev. D\textbf{75}, 108501 (2007)
[arXiv:gr-qc/0601020 [gr-qc]].

\bibitem{Holstein:2008sx}
B.~R.~Holstein and A.~Ross,
{\it Spin Effects in Long Range Gravitational Scattering},
[arXiv:0802.0716 [hep-ph]].

\bibitem{Anber:2011ut}
M.~M.~Anber and J.~F.~Donoghue,
{\it On the running of the gravitational constant},
Phys. Rev. D\textbf{85}, 104016 (2012)
[arXiv:1111.2875 [hep-th]].

\bibitem{Rodigast:2009zj}
A.~Rodigast and T.~Schuster,
{\it Gravitational Corrections to Yukawa and phi**4 Interactions},
Phys. Rev. Lett.\textbf{104}, 081301 (2010)
[arXiv:0908.2422 [hep-th]].

\bibitem{Marunovic:2012pr}
A.~Marunovic and T.~Prokopec,
{\it Antiscreening in perturbative quantum gravity and resolving the Newtonian singularity},
Phys. Rev. D\textbf{87}, no.10, 104027 (2013)
[arXiv:1209.4779 [hep-th]].

\bibitem{Bjerrum-Bohr:2015vda}
N.~E.~J.~Bjerrum-Bohr, J.~F.~Donoghue, B.~K.~El-Menoufi, B.~R.~Holstein, L.~Plant{\'e} and P.~Vanhove,
{\it The Equivalence Principle in a Quantum World},
Int. J. Mod. Phys. D\textbf{24}, no.12, 1544013 (2015)
[arXiv:1505.04974 [hep-th]].

\bibitem{Battista:2017xlm}
E.~Battista, A.~Tartaglia, G.~Esposito, D.~Lucchesi, M.~L.~Ruggiero, P.~Valko, S.~Dell'Agnello, L.~Di Fiore, J.~Simo and A.~Grado,
{\it Quantum time delay in the gravitational field of a rotating mass},
Class. Quant. Grav.\textbf{34}, no.16, 165008 (2017)
[arXiv:1703.08095 [gr-qc]].


 \bibitem{deBrito:2020wmp}
G.~P.~de Brito, M.~G.~Campos, L.~P.~R.~Ospedal and K.~P.~B.~Veiga,
{\it Quantum corrected gravitational potential beyond monopole-monopole interactions},
Phys. Rev. D\textbf{102}, no.8, 084015 (2020)
[arXiv:2006.12824 [hep-th]].

\bibitem{Akhoury:2013yua}
R.~Akhoury, R.~Saotome and G.~Sterman,
{\it High Energy Scattering in Perturbative Quantum Gravity at Next to Leading Power}
Phys. Rev. D\textbf{103}, no.6, 064036 (2021)
[arXiv:1308.5204 [hep-th]].

\bibitem{Saltas:2016nkg}
I.~D.~Saltas and V.~Vitagliano,
{\it Covariantly Quantum Galileon},
Phys. Rev. D\textbf{95}, no.10, 105002 (2017)
[arXiv:1611.07984 [hep-th]].


\bibitem{Saltas:2016awg}
I.~D.~Saltas and V.~Vitagliano,
{\it Quantum corrections for the cubic Galileon in the covariant language},
JCAP\textbf{05}, 020 (2017)
[arXiv:1612.08953 [hep-th]].

\bibitem{Latosh:2020jyq}
B.~Latosh,
{\it One-loop effective scalar-tensor gravity},
Eur. Phys. J. C \textbf{80}, no.9, 845 (2020)
[arXiv:2004.00927 [hep-th]].

\bibitem{Latosh:2022ydd}
B.~Latosh,
{\it FeynGrav: FeynCalc extension for gravity amplitudes},
Class. Quant. Grav. \textbf{39}, no.16, 165006 (2022)
[arXiv:2201.06812 [hep-th]].





\bibitem{vanDam:1970vg}
H.~van Dam and M.~J.~G.~Veltman,
{\it Massive and massless Yang-Mills and gravitational fields}
Nucl. Phys. B\textbf{22}, 397-411 (1970).

\bibitem{Bjerrum-Bohr:2014zsa}
N.~E.~J.~Bjerrum-Bohr, J.~F.~Donoghue, B.~R.~Holstein, L.~Plant{\'e} and P.~Vanhove,
{\it Bending of Light in Quantum Gravity},
Phys. Rev. Lett.\textbf{114}, no.6, 061301 (2015)
[arXiv:1410.7590 [hep-th]].

\bibitem{Jusufi:2016sym}
K.~Jusufi,
{\it Quantum effects on the deflection of light and the Gauss{\textendash}Bonnet theorem},
Int. J. Geom. Meth. Mod. Phys.\textbf{14}, no.10, 1750137 (2017)
[arXiv:1611.00713 [gr-qc]].



  
  
\bibitem{Goldberger:2004jt}
W.~D.~Goldberger and I.~Z.~Rothstein,
{\it An Effective field theory of gravity for extended objects}
Phys. Rev. D\textbf{73}, 104029 (2006)
[arXiv:hep-th/0409156 [hep-th]].

\bibitem{Foffa:2016rgu}
S.~Foffa, P.~Mastrolia, R.~Sturani and C.~Sturm,
{\it Effective field theory approach to the gravitational two-body dynamics, at fourth post-Newtonian order and quintic in the Newton constant},
Phys. Rev. D\textbf{95}, no.10, 104009 (2017)
[arXiv:1612.00482 [gr-qc]].

\bibitem{Levi:2018nxp}
M.~Levi,
{\it Effective Field Theories of Post-Newtonian Gravity: A comprehensive review},
Rept. Prog. Phys.\textbf{83}, no.7, 075901 (2020)
[arXiv:1807.01699 [hep-th]].  

\bibitem{Wu:2023jwd}
W.~H.~Wu and Y.~Tang,
{\it Post-Newtonian binary dynamics in the effective field theory of Horndeski gravity},
Chin. Phys. C\textbf{48}, no.3, 035101 (2024)
[arXiv:2312.02507 [gr-qc]].


\bibitem{Cheung:2018wkq}
C.~Cheung, I.~Z.~Rothstein and M.~P.~Solon,
{\it From Scattering Amplitudes to Classical Potentials in the Post-Minkowskian Expansion},
Phys.~Rev.~Lett.\textbf{121}, no.25, 251101 (2018)
[arXiv:1808.02489 [hep-th]].



\bibitem{Toms:2008dq}
D.~J.~Toms,
{\it Cosmological constant and quantum gravitational corrections to the running fine structure constant},
Phys. Rev. Lett.\textbf{101}, 131301 (2008)
[arXiv:0809.3897 [hep-th]].

\bibitem{Toms:2009zz}
D.~J.~Toms,
{\it Low energy quantum gravity, the cosmological constant and gauge coupling constants},
Int. J. Mod. Phys. D\textbf{17}, 2447 (2009).

\bibitem{Toms:2009vd}
D.~J.~Toms,
{\it Quantum gravity, gauge coupling constants, and the cosmological constant},
Phys. Rev. D\textbf{80}, 064040 (2009)
[arXiv:0908.3100 [hep-th]].

\bibitem{Toms:2010vy}
D.~J.~Toms,
{\it Quantum gravitational contributions to quantum electrodynamics},
Nature\textbf{468}, 56 (2010)
[arXiv:1010.0793 [hep-th]].

\bibitem{Toms:2011zza}
D.~J.~Toms,
{\it Quadratic divergences and quantum gravitational contributions to gauge coupling constants},
Phys. Rev. D\textbf{84}, 084016 (2011).

  


  
  \bibitem{Perez-Nadal:2008byr}
G.~Perez-Nadal, A.~Roura and E.~Verdaguer,
{\it Backreaction from non-conformal quantum fields in de Sitter spacetime},
Class. Quant. Grav.\textbf{25}, 154013 (2008)
[arXiv:0806.2634 [gr-qc]].


\bibitem{McDonald:2015iwt}
J.~I.~McDonald and G.~M.~Shore,
{\it Leptogenesis from loop effects in curved spacetime},
JHEP\textbf{04}, 030 (2016)
[arXiv:1512.02238 [hep-ph]].
  

\bibitem{Frob:2016fcr}
M.~B.~Fr{\"o}b and E.~Verdaguer,
{\it Quantum corrections to the gravitational potentials of a point source due to conformal fields in de Sitter},
JCAP\textbf{03}, 015 (2016)
[arXiv:1601.03561 [hep-th]].


\bibitem{Frob:2017smg}
M.~B.~Fr{\"o}b and E.~Verdaguer,
{\it Quantum corrections for spinning particles in de Sitter},
JCAP\textbf{04}, 022 (2017)
[arXiv:1701.06576 [hep-th]].

\bibitem{Boran:2017fsx}
S.~Boran, E.~O.~Kahya and S.~Park,
{\it Quantum gravity corrections to the conformally coupled scalar self-mass-squared on de Sitter background. II. Kinetic conformal cross terms},
Phys. Rev. D\textbf{96}, no.2, 025001 (2017)
[arXiv:1704.05880 [gr-qc]].  



\bibitem{Parker:2009uva}
L.~E.~Parker and D.~Toms,
{\it Quantum Field Theory in Curved Spacetime: Quantized Field and Gravity},
Cambridge University Press, 2009,
ISBN 978-0-521-87787-9, 978-0-521-87787-9, 978-0-511-60155-2
doi:10.1017/CBO9780511813924.

\bibitem{Buchbinder}
I.~L.~Buchbinder, S.~D.~Odintsov and I.~L.~Shapiro,
{\it Effective Action in Quantum Gravity},
Routledge, 2017,
ISBN 978-0-203-75892-2, 9780750301228, 978-0-7503-0122-0
doi:10.1201/9780203758922.


\bibitem{Bocharova:1970skc}
N.~M.~Bocharova, K.~A.~Bronnikov and V.~N.~Melnikov,
Vestn. Mosk.
Univ. Fiz. Astron. 6, 706 (1970).

\bibitem{Bekenstein:1974sf}
J.~D.~Bekenstein,
{\it Exact solutions of Einstein conformal scalar equations},
Annals Phys.\textbf{82}, 535-547 (1974).

\bibitem{Martinez:2002ru}
C.~Martinez, R.~Troncoso and J.~Zanelli,
{\it De Sitter black hole with a conformally coupled scalar field in four-dimensions},
Phys. Rev. D\textbf{67}, 024008 (2003)
[arXiv:hep-th/0205319 [hep-th]].


\bibitem{Bhattacharya:2013hvm}
S.~Bhattacharya and H.~Maeda,
{\it Can a black hole with conformal scalar hair rotate?}
Phys. Rev. D\textbf{89}, no.8, 087501 (2014)
[arXiv:1311.0087 [gr-qc]].


\bibitem{Moss:2014nya}
I.~G.~Moss,
{\it Covariant one-loop quantum gravity and Higgs inflation},
[arXiv:1409.2108 [hep-th]].  

\bibitem{Shapiro:2015ova}
I.~L.~Shapiro, P.~Morais Teixeira and A.~Wipf,
{\it On the functional renormalization group for the scalar field on curved background with non-minimal interaction},
Eur.~Phys.~J.~C\textbf{75}, 262 (2015)
[arXiv:1503.00874 [hep-th]].


\bibitem{Oda:2015sma}
K.~Y.~Oda and M.~Yamada,
{\it Non-minimal coupling in Higgs{\textendash}Yukawa model with asymptotically safe gravity},
Class.~Quant.~Grav.\textbf{33}, no.12, 125011 (2016)
[arXiv:1510.03734 [hep-th]].

\bibitem{Saltas:2015vsc}
I.~D.~Saltas,
{\it Higgs inflation and quantum gravity: An exact renormalisation group approach},
JCAP\textbf{02}, 048 (2016)
[arXiv:1512.06134 [hep-th]].
    

\bibitem{Bhattacharya:2017yix}
S.~Bhattacharya and T.~N.~Tomaras,
{\it Cosmic structure sizes in generic dark energy models},
Eur. Phys. J. C\textbf{77}, no.8, 526 (2017)
[arXiv:1703.07649 [gr-qc]].

\bibitem{Peskin:1995ev}
M.~E.~Peskin and D.~V.~Schroeder,
{\it An Introduction to quantum field theory},
Addison-Wesley, 1995,
ISBN 978-0-201-50397-5, 978-0-429-50355-9, 978-0-429-49417-8
doi:10.1201/9780429503559.



\bibitem{Seery:2007we}
D.~Seery,
{\it One-loop corrections to a scalar field during inflation},
JCAP\textbf{11}, 025 (2007)
[arXiv:0707.3377 [astro-ph]].  



\bibitem{Ivanov:2022qqt}
M.~M.~Ivanov and Z.~Zhou,
{\it Vanishing of Black Hole Tidal Love Numbers from Scattering Amplitudes},
Phys. Rev. Lett.\textbf{130}, no.9, 091403 (2023)
[arXiv:2209.14324 [hep-th]].  

 
  \bibitem{Clifton:2011jh}
T.~Clifton, P.~G.~Ferreira, A.~Padilla and C.~Skordis,
{\it Modified Gravity and Cosmology},
Phys. Rept. \textbf{513}, 1-189 (2012)
[arXiv:1106.2476 [astro-ph.CO]].
  

\end{thebibliography}
\end{document}